\documentclass[letterpaper]{article} 
\usepackage{aaai2026}  
\usepackage{times}  
\usepackage{helvet}  
\usepackage{courier}  
\usepackage[hyphens]{url}  
\usepackage{graphicx} 
\usepackage{natbib}  
\usepackage{caption} 
\usepackage{algorithm}
\usepackage{algorithmic}
\usepackage{amsmath} 
\usepackage{amsfonts}
\usepackage{multirow}
\usepackage{algorithmic}
\usepackage{algorithm}
\usepackage{xcolor}
\usepackage{colortbl, booktabs}
\usepackage{array}
\usepackage{makecell}
\usepackage{amssymb}
\usepackage{pifont}
\usepackage{array}
\usepackage{tikz}
\newcommand{\circnum}[2][\normalsize]{%
  \tikz[baseline=(char.base)]{
    \node[shape=circle,draw,inner sep=0.5pt] (char) {#1#2};}
}
\usepackage{newfloat}
\usepackage{listings}
\DeclareCaptionStyle{ruled}{labelfont=normalfont,labelsep=colon,strut=off} 
\floatstyle{ruled}
\newfloat{listing}{tb}{lst}{}
\floatname{listing}{Listing}
\title{Multi-Modal Style Transfer-based Prompt Tuning for Efficient 
Federated \\Domain Generalization}
\author{Yuliang Chen\textsuperscript{\rm 1,2}\equalcontrib, Xi Lin\textsuperscript{\rm 1,2}\equalcontrib, Jun Wu\textsuperscript{\rm 1,2}, Xiangrui Cai\textsuperscript{\rm 3}\thanks{Corresponding authors}, Qiaolun Zhang\textsuperscript{\rm 4}, \\Xichun Fan\textsuperscript{\rm 5}, Jiapeng Xu\textsuperscript{\rm 1,2}, Xiu Su\textsuperscript{\rm 6}\footnotemark[2]}

\affiliations{
    \textsuperscript{\rm 1}School of Computer Science, Shanghai Jiao Tong University\\
    \textsuperscript{\rm 2}Shanghai Key Laboratory of Integrated Administration Technologies for Information Security\\
    \textsuperscript{\rm 3}College of Computer Science, Nankai University\\
    \textsuperscript{\rm 4}Department of Electronics, Information and Bioengineering, Polytechnic Institute of Milan\\
    \textsuperscript{\rm 5}New York University Shanghai\\
    \textsuperscript{\rm 6}Big Data Institute, Central South University\\

    \{chenyuliang, linxi234, junwuhn, xjp20021018\}@sjtu.edu.cn, caixr@nankai.edu.cn, qiaolun.zhang@mail.polimi.it, xf731@nyu.edu, xiusu1994@csu.edu.cn
}

\usepackage{bibentry}

\begin{document}

\maketitle

\begin{abstract}
Federated Domain Generalization (FDG) aims to collaboratively train a global model across distributed clients that can generalize well on unseen domains.
However, existing FDG methods typically struggle with cross-client data heterogeneity and incur significant communication and computation overhead. To address these challenges, this paper presents a new FDG framework, dubbed FaST-PT, which facilitates local feature augmentation and efficient unseen domain adaptation in a distributed manner. First, we propose a lightweight Multi-Modal Style Transfer (MST) method to transform image embedding under text supervision, which could expand the training data distribution and mitigate domain shift. 
We then design a dual-prompt module that decomposes the prompt into global and domain prompts. 
Specifically, global prompts capture general knowledge from augmented embedding across clients, while domain prompts capture domain-specific knowledge from local data.
Besides, Domain-aware Prompt Generation (DPG) is introduced to adaptively generate suitable prompts for each sample, which facilitates unseen domain adaptation through knowledge fusion.
Extensive experiments on four cross-domain benchmark datasets, e.g., PACS and DomainNet, demonstrate the superior performance of FaST-PT over SOTA FDG methods such as FedDG-GA and DiPrompt. Ablation studies further validate the effectiveness and efficiency of FaST-PT.
\end{abstract}


\section{Introduction}

Federated Learning (FL) \cite{mcmahan2017communication} is a widely adopted distributed machine learning paradigm, which facilitates collaborative model training while protecting privacy.
Instead of relying on a single centralized server, FL distributes the training process across multiple clients (e.g., mobile devices or institutions) to collaboratively train a global model without sharing their private data.
It is specifically well-suited for real-world applications where sensitive data is distributed across multiple sources \cite{guan2024federated}.

Although FL has shown promising performance in various tasks, existing studies predominantly focus on enhancing model performance for the participating clients.
However, they often overlook the generalization capability to unseen domains beyond the participating clients. 
Data encountered in these unseen environments can exhibit substantial distributional differences from training clients, due to variations in data sources, collection methods, or domain-specific characteristics. 
Effectively adapting FL to such domain shifts \cite{muandet2013domain,li2018domain} remains a technically challenging problem. 

Federated Domain Generalization is defined as training a global model on multiple distributed source domains, aiming to achieve robust generalization to completely unseen target domains.
Early attempts have been made to address the FDG problem. For example,  ELCFS \cite{liu2021feddg} allows clients to share the amplitude spectrum in frequency space for federated medical image segmentation.
Similarly, CCST \cite{chen2023federated} extracts and exchanges style information to perform style transfer. Besides, FPL \cite{huang2023rethinking} facilitates the exchange of prototypes among clients. A common limitation of these methods is their reliance on sharing domain-related information among clients, which compromises the data privacy and communication constraints inherent in FL. 
Additionally, extracting domain-specific information often requires auxiliary modules, leading to increased computational overhead. 

To address the above challenges, we propose a new framework dubbed \underline{F}ederated  Multi-Mod\underline{a}l \underline{S}tyle \underline{T}ransfer with \underline{P}rompt \underline{T}uning (FaST-PT). In FaST-PT, we introduce CLIP into FL and utilize its text-image alignment and generalization ability for handling FDG problems.
On the one hand, we leverage CLIP's powerful image-text alignment capabilities and propose Multi-Modal Style Transfer (MST), which enables client-side external domain style transfer to expand the training data distribution. 
Specifically, MST introduces a transform network that transforms image embeddings using the text representation of the external domain, facilitating the style transfer while retaining the class-specific information of the original image.  
On the other hand, we employ the prompt tuning paradigm of CLIP to facilitate efficient unseen domain adaptation.
 We first design a dual-prompt module that incorporates both global and domain prompts.
 In detail, global prompts are trained on augmented embeddings and aggregated across clients to capture general knowledge, while local prompts are trained solely on local data to preserve domain-specific knowledge.
Additionally, we propose a domain-aware prompt generation mechanism to further incorporate domain knowledge based on domain similarity, thereby enhancing the adaptability of prompts to unseen domains.
Our contribution can be summarized as follows:
\begin{itemize}
    \item We present FaST-PT, a new FL framework that tackles the FDG problem through both client-side distribution alignment and efficient unseen domain adaptation by leveraging the text-image alignment and generalization capabilities of the pre-trained CLIP model.
    \item We propose a multi-modal style transfer method that allows clients to transform the image embedding to the style of the external domains while retaining the class information under the supervision of text description to mitigate the distribution discrepancy among clients.
    \item We design a dual-prompt module that leverages prompt partition to simplify feature disentanglement in traditional domain generalization, and employs domain-aware prompt generation to integrate domain knowledge for efficient unseen domain adaptation.
    \item Extensive experiments are conducted with FaST-PT on four benchmark datasets, consistently demonstrating its superior performance over baseline methods.
Comprehensive ablation studies further validate the effectiveness and efficiency of FaST-PT.
\end{itemize}

\begin{figure*}
    \centering
    \includegraphics[width=1\linewidth]{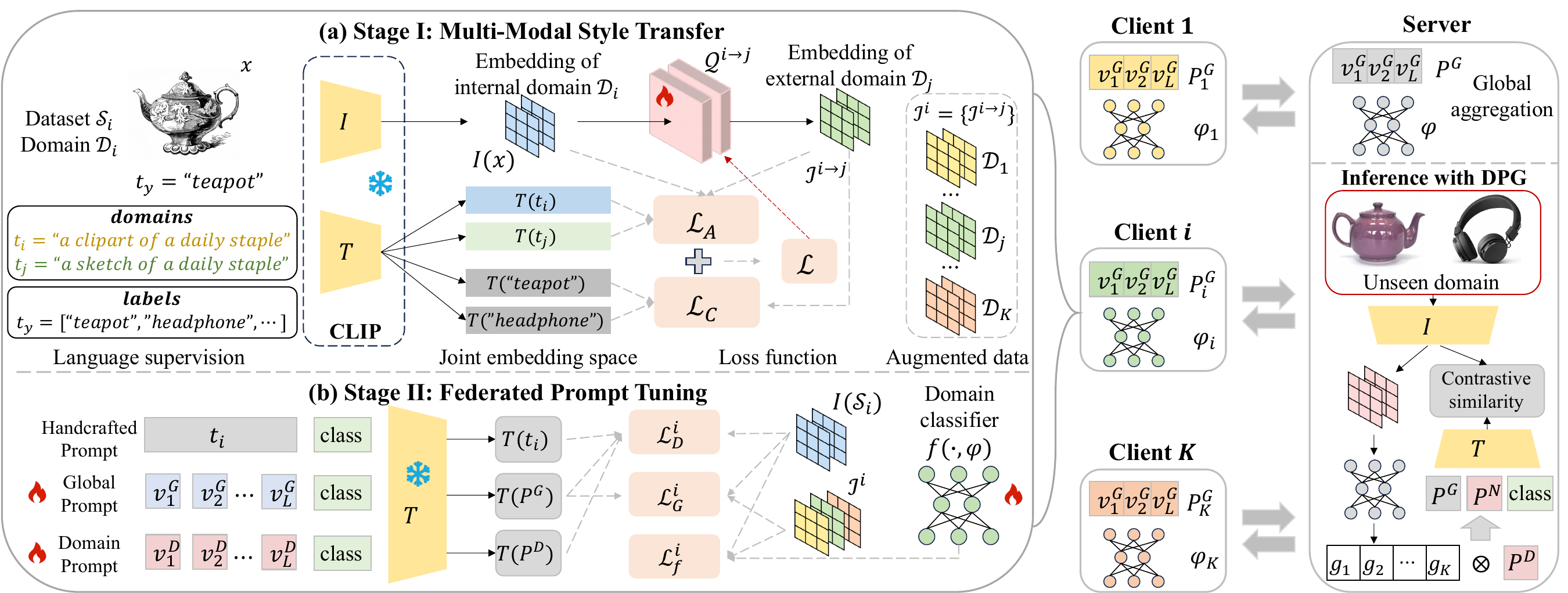}
    \caption{Framework of our proposed FaST-PT. The left side illustrates the local training process, while the right side depicts the communication between client and server. The client-side training is divided into two phases: (a) MST and (b) FPT.
(a) In the MST phase, the client employs a frozen CLIP model along with text descriptions from various domains to train a transform network, enabling style transfer at the embedding level.
(b) In the FPT phase, the client trains global prompts, domain prompts, and domain classifiers using the augmented data.
The inference process is also depicted on the server side.}
    \label{fig:frame}
\end{figure*}

\section{Related Work}
\subsection{Federated Learning}
Federated learning (FL) distributes the training progress across multiple devices or compute nodes instead of a single central server to facilitate global training of the model while protecting data privacy. The most typical FL framework is FedAvg \cite{mcmahan2017communication,konevcny2016federated}, which decomposes the distributed learning process from the perspectives of local training and global averaging. However, traditional FL performs poorly under domain shift. 
Existing methods mainly focus on regularizing the local training on the client side to attain a well-performed global model.
FedProx \cite{li2020federated} added an extra term in the loss function to control the $L_2$ distance of model weights.
FedBN \cite{li2021fedbn} retains the parameters of the batch normalization layers locally and excludes them from the aggregation process.
\subsection{Federated Domain Generalization}

Federated domain generalization is a variant of DG tailored to the FL setting, aiming to mitigate domain shifts across clients and improve the model's generalization to unseen domains \cite{liu2021feddg,nguyen2022fedsr,wu2021collaborative,chen2023federated,zhang2021federated,huang2023rethinking,le2024efficiently}. 
 ELCFS \cite{liu2021feddg} is the first to solve the FDG problem, which exchanges the amplitude spectrum in the frequency domain and employs episodic learning to further enhance generalization performance.
CCST \cite{chen2023federated} allows clients to share their style information and achieve local style transfer, leading to more uniform distributions of source clients.
Apart from domain information sharing methods, the remaining approaches focus on extracting domain-invariant features.
For example, FedSR \cite{nguyen2022fedsr} leverages L2-norm and conditional mutual information regularizations to learn simple yet generalizable data representations.
COPA \cite{wu2021collaborative} simultaneously incorporates a domain-invariant representation extractor along with an ensemble of domain-specific classifiers.

\subsection{Prompt Tuning}
Prompt tuning \cite{lester2021power} is a parameter-efficient fine-tuning technique that introduces a small set of learnable prompt tokens as additional inputs, while keeping the parameters of foundation models frozen. As a pioneering work, CoOp \cite{zhou2022learning} adapts CLIP \cite{radford2021learning} to downstream image recognition tasks by converting the input context tokens in the text encoder into learnable prompts. CoCoOp \cite{zhou2022conditional} extends CoOp by generating input-dependent prompts through a lightweight network. MaPLe \cite{khattak2023maple} further enhances the alignment between textual and visual representations in CLIP by introducing visual prompts alongside textual ones.
In recent years, prompt tuning has been introduced into federated learning due to its inherent suitability for efficient communication. For instance, PromptFL \cite{guo2023promptfl} and FedPrompt \cite{zhao2023fedprompt} independently learn text prompts on each local client and aggregate the optimized prompts on the server.

\section{Methods}
We assume $K$ clients, where each client $i$ possesses a dataset from domain $\mathcal{D}_i$.
The primary objective is to jointly train a global prompt module that generalizes effectively to the unseen target domain $\mathcal{D}_u$.
A key assumption in our framework is that each client has access to simple text descriptions of external training domains $t_i$ (e.g., ``a clipart of a'' ), which are considered non-sensitive and do not compromise client privacy.
The framework of FaST-PT is shown in Figure \ref{fig:frame}. 

\subsection{Multi-Modal Style Transfer}
Existing FDG approaches improve the generalization capability of local models by incorporating knowledge from diverse domains, which necessitates sharing part of the domain knowledge among clients. While sharing such sensitive information between clients may bring the risk of data leakage, we propose a lightweight and secure local style transfer scheme, MST, which leverages CLIP's extensive domain knowledge and strong text-image alignment capabilities to facilitate latent feature augmentation at the embedding level. It allows client $k$ with domain $\mathcal{D}_k$ to generate image embeddings for external domains $\mathcal{D}_{\{1,2\dots K\}\backslash k}$.

MST mainly learns a transform network $\mathcal{Q}^{i\rightarrow j}$ that transforms the image embedding of the input image from $\mathcal{D}_i$ to $\mathcal{D}_j$, with the goal formulated as the following two aspects: 

\begin{itemize}
    \item The alignment between $\mathcal{D}_i$ and $\mathcal{D}_j$
    \item The preservation of class-consistent features
\end{itemize}
Let $x$ denote the input image and $I(x)$ denote the image embedding of $x$. Besides, let $t_k$ denote the domain-related description of domain $\mathcal{D}_k$ and $t_y$ denote the class label. We compose these two parts to obtain the text input as $t = [t_k,t_y]$, which is further encoded to get the text embedding $T(t)$. For instance,  if $t_i=$ ``a clipart of a \{\}'', $t_j=$ ``a sketch of a \{\}'', and $t_y$ could be ``tiger'', then the composed text $t=$``a clipart of a tiger''.

To achieve the embedding transform from $\mathcal{D}_i$ to $\mathcal{D}_j$ while preserving class-consistent features, we design two loss functions to train $\mathcal{Q}^{i\rightarrow j}$ for each domain $\mathcal{D}_j$ distinct from the local domain $\mathcal{D}_i$: alignment loss and consistency loss.
\paragraph{Alignment Loss} The alignment loss $\mathcal{L}_{\text{A}}(\mathcal{Q}^{i\rightarrow j})$ aims to ensure that the embeddings of augmented images more faithfully reflect the distinctive properties of their respective domains. $\mathcal{L}_{\text{A}}(\mathcal{Q}^{i\rightarrow j})$ is supervised by textual descriptions associated with different source domains. While CLIP effectively captures relationships between text and image embeddings, directly obtaining their precise correspondence remains challenging. To address this, we align the change direction in the image and text embedding spaces as the optimization target. Specifically, this change direction is defined as the normalized difference between embeddings from $\mathcal{D}_i$ and $\mathcal{D}_j$. The alignment loss is thus formulated as:

\begin{equation}
\begin{split}
\mathcal{L}_{\text{A}}(\mathcal{Q}^{i\rightarrow j}) = \sum_{(x,y)\in\mathcal{S}_i} &\left( 1 -  \frac{\mathcal{Q}^{i\rightarrow j}(I(x)) - I(x)}{\|\mathcal{Q}^{i\rightarrow j}(I(x)) - I(x)\|} \right. \\
&\left. \cdot \frac{T(t_{j}, t_y) - T(t_{i}, t_y)}{\|T(t_{j}, t_y) - T(t_{i}, t_y)\|} \right).
\end{split}
\end{equation}

\paragraph{Consistency Loss} The consistency loss $\mathcal{L}_{\text{C}}$ is designed to preserve the class knowledge in transformed embeddings. The intuition behind designing $\mathcal{L}_{\text{C}}$ is that using only the alignment loss may ensure style transfer of the embedding, but it can also erase some class information. We leverage CLIP’s zero-shot inference capabilities to assess the class information contained in the enhanced image embeddings. Specifically, we perform classification by comparing these enhanced embeddings with the text embeddings of class names. To this end, we adopt CLIP’s standard contrastive loss, formulated as follows:
\begin{equation}
    \mathcal{L}_{\text{C}}(\mathcal{Q}^{i\rightarrow j}) = \mathbb{E}_{(x,y)\in\mathcal{S}_i} \ell_{ce} ([\mathcal{Q}^{i\rightarrow j}(I(x)) \cdot T(t_{y})], y),
\end{equation}
where $\left[I \cdot T\right]$ denote the Softmax-based contrastive similarity computation and Top-1 prediction process.

The general objective $\mathcal{L}_{\text{MST}}(\mathcal{Q}^{i\rightarrow j})$ of training the transform network $\mathcal{Q}^{i\rightarrow j}$ is the composition of the above two loss functions. To balance the effectiveness of alignment and consistency, we additionally introduce a hyperparameter $\lambda$ to achieve the linear composition: 
\begin{equation}\label{loss}
\mathcal{L}_{\text{MST}}(\mathcal{Q}^{i\rightarrow j}) = \lambda \mathcal{L}_{\text{A}}(\mathcal{Q}^{i\rightarrow j}) + (1-\lambda)\mathcal{L}_{\text{C}}(\mathcal{Q}^{i\rightarrow j}).
\end{equation}

Client $i$ with domain $\mathcal{D}_i$ locally learn $(K-1)$ transform networks for the other $(K-1)$ clients and generate the corresponding augmentation embeddings $\mathcal{I}^{i\rightarrow j} = \mathcal{Q}^{i\rightarrow j}(x), x\in \mathcal{S}_i$. Thus, each client maintains a set of augmentation embeddings as $ \mathcal{I}_{i} = \{ \mathcal{I}^{i\rightarrow j} \}, j \in \{1,2\dots K\}\backslash i$.

\subsection{Federated Prompt Tuning}
Yet, conventional DG methods often involve complex feature extraction techniques. How to leverage augmented embeddings to achieve domain generalization with minimal resource consumption remains an open challenge.
Building on the powerful feature extraction capabilities of CLIP, we introduce prompt tuning of CLIP for efficient domain generalization. Motivated by the idea of feature disentanglement \cite{bui2021exploiting,zhang2022towards} in traditional DG, we propose a dual-prompt module that divides the prompts into global and domain prompts.

\paragraph{Global Prompt} $P^G$ is designed to capture the general knowledge of the downstream task, which should be shared across clients. $P^G$ is a set of $L$ learnable vectors $P^G=\{v^G_1,v^G_2,\dots,v^G_L\}$, with the same dimension as text embedding $d$. We denote the class label as $t_y$. For the $j$-th class, we define the input of the text encoder with global prompt as $t = [P^G,t_y^j]$. We denote the feature map of image $x$ as $I(x)$ and the text feature map as $T([P^G,t_y^j])$. Thus, the prediction probability of the $j$-th class computed with contrastive similarity can be formulated as:
\begin{equation}
    p_G(y=j|x)=\frac{exp(\mathcal{H}(T([P^G,t_y^j]),I(x))/\tau)}{\sum_{i=1}^{C} exp(\mathcal{H}(T([P^G,t_y^i]),I(x))/\tau)},
\end{equation}
where $\mathcal{H}$ is the cosine similarity, $C$ is the number of class and $\tau$ is the temperature of $\text{Softmax}$ function. We optimize the global prompts by minimizing the cross-entropy loss computed over both the augmented embeddings and the original dataset. Specifically, the training loss on client $i$ is defined as follows:
\begin{equation}\label{1}
    \mathcal{L}_\text{G}^i = \mathbb{E}_{(x,y)\in\mathcal{S}_i\cup \mathcal{I}_i}[\ell_{ce}(y, p_{G}(y|x))].
\end{equation}
\paragraph{Domain Prompt} $P^D_i$ is designed to capture the domain-specific knowledge of client $i$. Similarly, $P^D_i$ is a set of vectors $P^D_i=\{v^D_1,v^D_2,\dots,v^D_L\}$ with length $L$ and dimension $d$. We concatenate $P^G$ with the domain prompt to form a new composite prompt $P^C = [P^G,P^D_i]$. Specifically, the predicted probability is updated accordingly after introducing the domain prompt, by replacing $P^G$ with $P^C$. The core idea of the domain-specific prompt is to enhance the performance of $P^D_i$ on local data $\mathcal{S}_i$. We similarly optimize it using the cross-entropy loss $\mathcal{L}_{\text{Cla}}^i=\mathbb{E}_{(x,y)\in\mathcal{S}_i}[\ell_{ce}(y, p_{C}(y|x))]$.
Besides, the training of $P^D_i$ should also enforce orthogonality with $P^G$ to guarantee the uniqueness of the extracted knowledge. To this end, we additionally introduce a contrastive loss term as follows:
\begin{equation}
\begin{split}
    \mathcal{L}_\text{D}^i  =\mathcal{L}_{\text{Cla}}^i+\mathcal{L}_{\text{Con}}^i = \mathbb{E}_{(x,y)\in\mathcal{S}_i}[\ell_{ce}(y, p_{C}(y|x))]\\
    -\text{log}\frac{exp(sim(P^D_i,t_i))}{exp(sim(P^D_i,t_i))+exp(sim(P^D_i,P^G))},
    \end{split}
\end{equation}
where $t_i$ is the handcrafted prompt for Domain $\mathcal{D}_i$ and $p_{C}(y|x)$ is the probability predicted with prompt $P^C$.

Notably, clients share global prompts for aggregation during training, while domain-specific prompts are retained locally and only shared after training. The final domain prompt list is formed as $P^D=[P^D_1, P^D_2, \dots, P^D_K]$.

\begin{algorithm}[!h]
    \begin{algorithmic}[1]
        \STATE{\bfseries Input:} Datasets $\mathcal{S}_i$, client number $K$, training round $R$, epoch number $E_G$ and $E_L$
        \STATE{\bfseries Output:} Global Prompt $P^G$, Domain prompt list $P^D$, domain classifier $f(\cdot,\varphi)$
        \STATE{\bfseries Stage 1: Multi-Modal Style Transfer} 
        \FOR{client $i=1,2,\ldots, K$}
            \FOR{Domain $\mathcal{D}_j \in \{\mathcal{D}_1,\mathcal{D}_2,\ldots, \mathcal{D}_K\} \backslash \mathcal{D}_i $ }
            \STATE Train $\mathcal{Q}^{i\rightarrow j}$  with $\mathcal{L}_{\text{MST}}(\mathcal{Q}^{i\rightarrow j})$ (Eq. \ref{loss})
            \ENDFOR
        \ENDFOR
        \STATE Clients locally augments image embedding $\mathcal{I}^{i\rightarrow j}$
        \STATE{\bfseries Stage 2: Federated Prompt Tuning} 
        \FOR{round $r=1,2,\ldots, R$}
        \FOR{client $i=1,2,\ldots, K$} 
        \FOR{epoch $e=1,2,\ldots,E_G$}
        \STATE Training $P^G_i$ and $\varphi_i$ with $\mathcal{L}_\text{G}^i$ and $\mathcal{L}_\text{F}^i$
        \ENDFOR
        \ENDFOR
        \STATE Upload $P^G_i$ and $\varphi_i$
        \STATE Server aggregates $P^G$ and $\varphi$ and broadcasts
        \FOR{client $i=1,2,\ldots, K$} 
        \FOR{epoch $e=1,2,\ldots,E_D$}
        \STATE Training $P^D_i$ with $\mathcal{L}_\text{D}^i$ 
        \ENDFOR
        \ENDFOR
        \ENDFOR
        \STATE Server collects $P^D$ and broadcasts
    \end{algorithmic}
    \caption{FaST-PT}
    \label{alg1}

\end{algorithm}

\paragraph{Domain-aware Prompt Generation}
Considering the latent similarity between training and unseen domains, we further design a domain-aware prompt generation scheme to integrate knowledge from diverse domains to enhance generalization performance, which enables better adaptation of the target domain $\mathcal{D}_u$. We utilize a domain classifier $f(\cdot,\varphi)$ to facilitate the domain similarity analysis. $f(\cdot,\varphi)$ takes image feature as the input and outputs the one-hot format result with $K$ dimension as: $ \mathbf{G} = \begin{bmatrix} g_1,g_2 \dots g_K \end{bmatrix} $.

\begin{table*}[tb]
  \centering
  \setlength{\tabcolsep}{10pt}
  \renewcommand{\arraystretch}{1}
  {\small
  \begin{tabular}{l||ccccc||ccccc}
    \hline\hline
    \rowcolor{gray!20}
      & \multicolumn{5}{c||}{\textbf{PACS (\%)}} & \multicolumn{5}{c}{\textbf{VLCS (\%)}} \\
    \cline{2-11} 
\rowcolor{gray!20} 
\multirow{-2}{*}{Methods} 
   & A & C & P & S & Avg & C & L & S & V & Avg \\
    \hline
    \hline

    FedAvg &81.27&76.78&92.33&63.54&78.48&92.78&62.39&73.02&74.98&75.79\\

    \rowcolor{gray!10}
    FedProx &80.22&77.17&96.08&64.28&79.44&93.22&62.80&74.64&74.96&76.41\\
    \hline
     ELCFS &85.89&74.02&92.14&69.16&80.30&93.28&61.64&73.42&76.06&76.10\\
    \rowcolor{gray!10} FedSR &84.97&85.13&91.37&67.13&82.15&95.55&63.29&71.93&75.43&76.55\\
     FedADG &80.16&81.92&91.28&68.43&80.45&95.96&65.63&70.29&76.70&77.15\\
    \rowcolor{gray!10}  FedDG-GA &94.03&83.19&76.85&82.93&84.25&95.14&62.82&74.41&77.49&77.47\\
    \hline
      PromptFL &93.19&94.72&99.10&83.76&92.70&99.26&66.08&79.38&78.56&80.82\\
        \rowcolor{gray!10}  FedAPT&93.78&95.52&99.63&85.31&93.56&99.01&67.91&78.66&81.53&81.78\\
        DiPrompt &95.72&96.39&99.24&88.07&94.86&\textbf{99.70}&68.97&84.72&82.93&84.08\\
        
    \hline 
    \rowcolor{gray!10}
    \textit{FaST-PT} &\textbf{97.32}&\textbf{99.31}&\textbf{99.93}&\textbf{96.09}&\textbf{98.16}&99.17&\textbf{70.59}&\textbf{86.05}&\textbf{86.26}&\textbf{85.52}\\
    \hline\hline
  \end{tabular}}
  
   \caption{Performance comparison of different methods on the PACS dataset and the VLCS dataset.} 

   \label{tab:0}

\end{table*}

We weight $\otimes$ the domain prompt list $P^D$ with the pseudo-domain label $ \mathbf{G}$ element-wise to generate the dynamic prompt $P^N$ for arbitrary image $x$ as follows:
\begin{equation}
    P^N = f(I(x),\varphi) \otimes P^D.
\end{equation}

$P^N$ integrates knowledge from multiple source domains to generate knowledge that approximates the target domain. During the inference stage, $P^N$ is concatenated with $P^G$ and fed into the text encoder. For the optimization of $f(\cdot,\varphi)$, we leverage the augmented dataset $\mathcal{S}_i \cup \mathcal{I}_i$ for training and denote the domain index as the label. We utilize the cross-entropy loss and formulate the objective as follows:

\begin{equation}\label{2}
    \mathcal{L}_\text{F}^i = \mathbb{E}_{(x,y,j)\in \mathcal{S}_i \cup \mathcal{I}_i}[\ell_{ce}(j,f(x,\varphi_i))],
\end{equation}
where $j$ is the domain index.

During the training of the prompt module, we adopt a two-stage alternating update strategy. First, clients update the global parameters $P^G$ and $\varphi$, and upload the gradient updates, which are then aggregated by the server and broadcast back to all clients. Second, clients update the local parameters $P^D_i$ and upload them after the final round. Since the training of $P^G$ and $f(\cdot,\varphi)$ does not interfere with each other, it is reasonable to train them simultaneously. The complete training process is illustrated in Alg.\ref{alg1}.

\subsection{Discussion}
FaST-PT achieves efficient local feature enhancement and robust domain generalization. Specifically, although FaST-PT leverages MST to mitigate domain shift and adopts prompt tuning for domain generalization, both components are built upon the CLIP model. This unified backbone design obviates the need for deploying additional models on local clients, thereby reducing memory consumption and computational overhead. Moreover, unlike traditional federated learning methods that require sharing full model parameters, FaST-PT only exchanges prompts and their associated lightweight generation modules across clients, substantially lowering communication costs (as further evidenced in our experimental results). Finally, in contrast to approaches that transmit domain-specific features, MST merely requires clients to share text descriptions of their respective domains—information that is inherently less sensitive, thus providing stronger privacy protection.

\section{Experiments}
\subsection{Experimental Setup}
\paragraph{Datasets} We perform extensive experiments on four cross-domain datasets: 
\textbf{PACS} \cite{li2017deeper} has four domains, \textit{photo} (P), \textit{art-painting} (A), \textit{cartoon} (C) and \textit{sketch} (S). 
\textbf{VLCS} \cite{fang2013unbiased} is a widely used benchmark for domain generalization, consisting of four domains, \textit{VOC2007} 
 (V), \textit{LabelMe} (L), \textit{Caltech101} (C), and \textit{SUN09} (S). 
\textbf{DomainNet} \cite{peng2019moment} is a large-scale dataset across six domains, \textit{clipart} (C), \textit{infograph} (I), \textit{painting} (P), \textit{quickdraw} (Q), \textit{real} (R), and \textit{sketch} (S). 
\textbf{OfficeHome} \cite{venkateswara2017deep} is a medium-scale benchmark dataset containing four domains: \textit{Art} (A), \textit{Clipart} (C), \textit{Product} (P), and \textit{Real-World} (R). 
Following \cite{li2021fedbn,yang2023efficient}, we select the top 10 classes for our experiment on DomainNet and OfficeHome.
We adopt the leave-one-domain-out protocol, treating one domain as the target while using the others as source domains.

\begin{table*}[tb]

  \centering
  \setlength{\tabcolsep}{6.5pt}
  \renewcommand{\arraystretch}{1}
    {\small
  \begin{tabular}{l||ccccc||ccccccc}
    \hline\hline
    \rowcolor{gray!20}
     & \multicolumn{5}{c||}{\textbf{OfficeHome (\%)}} & \multicolumn{7}{c}{\textbf{DomainNet (\%)}} \\
        \cline{2-13} 
\rowcolor{gray!20} 
\multirow{-2}{*}{Methods}& A & C & P & R & Avg & C & I & P & Q & R & S & Avg \\
    \hline
    \hline

    FedAvg &82.56&79.20&88.28&91.15&85.30&78.04&62.69&80.13&45.78&82.49&84.17&72.22\\

    \rowcolor{gray!10}
    FedProx &83.19&77.91&90.28&91.60&85.74&79.11&64.77&82.82&51.65&84.65&83.94&74.49\\
    \hline
     ELCFS &88.72&82.64&89.37&91.75&88.12&83.28&67.18&85.78&58.65&88.31&86.37&78.26\\

    \rowcolor{gray!10} FedSR &88.17&84.47&88.89&90.65&88.04&85.17&66.57&86.01&60.95&87.26&90.10&79.34\\
     FedADG &86.51&87.25&91.52&92.30&89.40&83.95&69.68&84.05&61.51&90.78&86.79&79.46\\
    \rowcolor{gray!10} FedDG-GA &87.86&86.90&92.42&93.25&90.11&86.69&70.93&85.69&60.72&89.82&87.18&80.17\\
    \hline
     PromptFL  &92.35&88.73&95.20&98.80&93.77&89.92 &72.75& 87.46&60.85 & 93.54& 90.47&82.50\\
        \rowcolor{gray!10} FedAPT &94.28&90.76&97.08&98.95&95.27&91.44&72.04&90.62&63.88&95.27&94.23&84.58\\
        DiPrompt&95.48&91.26&97.90&99.10&95.94&93.89&75.34&93.72&64.16&96.31&97.62&86.84\\
    \hline 
    \rowcolor{gray!10}
    \textit{FaST-PT} &\textbf{98.57}&\textbf{92.72}&\textbf{99.24}&\textbf{99.40}&\textbf{97.48}&\textbf{97.48}&\textbf{78.14}&\textbf{97.17}&\textbf{66.88}&\textbf{97.90}&\textbf{98.46}&\textbf{89.17}\\
    \hline\hline
  \end{tabular}}
  
    \caption{Performance comparison of different methods on the OfficeHome dataset and the DomainNet dataset. }
      \label{tab:1}
   
\end{table*}

\paragraph{Baselines}
We compare our method with three types of state-of-the-art methods: 1) Traditional federated learning framework including \textbf{FedAvg} \cite{mcmahan2017communication} and \textbf{FedProx} \cite{li2020federated} that enables clients to collaboratively train a global model from scratch. 
2) Conventional FedDG methods that integrate DG with FL to collaboratively train a generalized model that perform well on unseen domains, including \textbf{ELCFS} \cite{liu2021feddg}, \textbf{FedSR} \cite{nguyen2022fedsr}, \textbf{FedADG} \cite{zhang2021federated2} and \textbf{FedDG-GA} \cite{zhang2021federated}.
3) Prompt tuning-based methods that introduce prompt tuning into FL to enhance the generalization ability of CLIP, including
\textbf{PromptFL} \cite{guo2023promptfl}, \textbf{FedAPT} \cite{su2024federated}, and \textbf{DiPrompt} \cite{bai2024diprompt}.

\paragraph{Implement Details}
We conduct our experiment with Python 3.11 on NVIDIA RTX A40 GPU. We use the OpenAI CLIP model with a ViT-L/14 backbone, which encodes both images and text into 768-dimensional feature vectors. For the MST stage, we utilize a 2-layer MLP as the transform network $\mathcal{Q}^{{i\rightarrow j}}$ with a hidden dimension of 384. we utilize the Adam optimizer with a learning rate of $1e^{-3}$ and a weight decay of 0.05. The loss weight $\lambda$ is set as $0.5$. For the prompt training stage, we utilize a fully-connection layer as the domain classifier $f(\cdot,\varphi)$. We use the SGD optimizer with a learning rate of $0.005$ and $0.01$ for training of the prompt and domain classifier, respectively. We set the communication round $R$ as 20 and local training epoch $E_G=E_L=5$. We set the batch size as 16 and prompt length as 4. We use the same hyperparameters and backbones across all experiments to ensure a fair comparison.

\subsection{Evaluation}
We demonstrate the experimental results of FaST-PT and other methods in Table \ref{tab:0} and Table \ref{tab:1}, which consists of totally 18 domain generalization tasks on PACS, VLCS, OfficeHome and DomainNet. 
 All experimental results are reported as the average over three runs with different random seeds. 
Generally, we observe that although conventional FDG methods have demonstrated significantly better generalization performance than traditional FL approaches, there remains a notable performance gap compared to recent prompt-based FL methods. This substantial improvement in generalization can be largely attributed to the strong cross-modal and generalization capabilities of CLIP. Even when all methods are built upon the same pretrained visual backbone (ViT-L/14), prompt tuning enables more efficient adaptation to the distributional characteristics of each domain.
Specifically, FaST-PT achieves state-of-the-art performance with an average accuracy of 98.16\% on PACS, 85.52\% on VLCS, 97.48\% on OfficeHome, and 89.17\% on DomainNet, surpassing the SOTA method DiPrompt by up to +3.30\% on PACS and +2.33\% on DomainNet. Among the total of 18 generalization tasks across all benchmarks, FaST-PT achieves the best performance on 17 tasks, demonstrating remarkable consistency. 
Compared to existing prompt-based FL methods, FaST-PT still demonstrates a substantial performance advantage, highlighting the effectiveness of the proposed MST and DAP strategy, both of which play a critical role in enhancing cross-domain adaptability and mitigating the adverse effects of distributional shifts in FL.

\subsection{Ablation Study}
\paragraph{Effectiveness of Components}
To investigate the effectiveness of different components in FaST-PT, we evaluated various variants of FaST-PT on the PACS dataset, as shown in Table \ref{tab:ablation}.
In variants \circnum[\scriptsize]{1} and \circnum[\scriptsize]{2}, we only employ global prompts or domain prompts combined with the DPG mechanism. Compared to variant \circnum[\scriptsize]{4}, the performance decreased noticeably, which demonstrates the effectiveness of integrating global and domain knowledge for generalization tasks.
 In addition, through two comparative analyzes ( \circnum[\scriptsize]{1} vs.  \circnum[\scriptsize]{3} and \circnum[\scriptsize]{4} vs. FaST-PT), we observed that the introduction of the MST strategy can mitigate the domain shift problem in FDG and effectively improve model performance, thus improving the generalization capability of the global model. 
Variant \circnum[\scriptsize]{5} exhibits a performance decline compared to FaST-PT, which also reflects that the contrastive loss we employed can strengthen the separation between global and domain knowledge, thereby indirectly enhancing adaptation to the target domain.
In Variant \circnum[\scriptsize]{6}, we conducted experiments by supplementing the textual descriptions of the target domain, allowing clients to generate image features for the target domain to assist the training of the global model. Its performance only surpasses FaST-PT by 0.24\%. 

\begin{figure}
    \centering
    \includegraphics[width=1\linewidth]{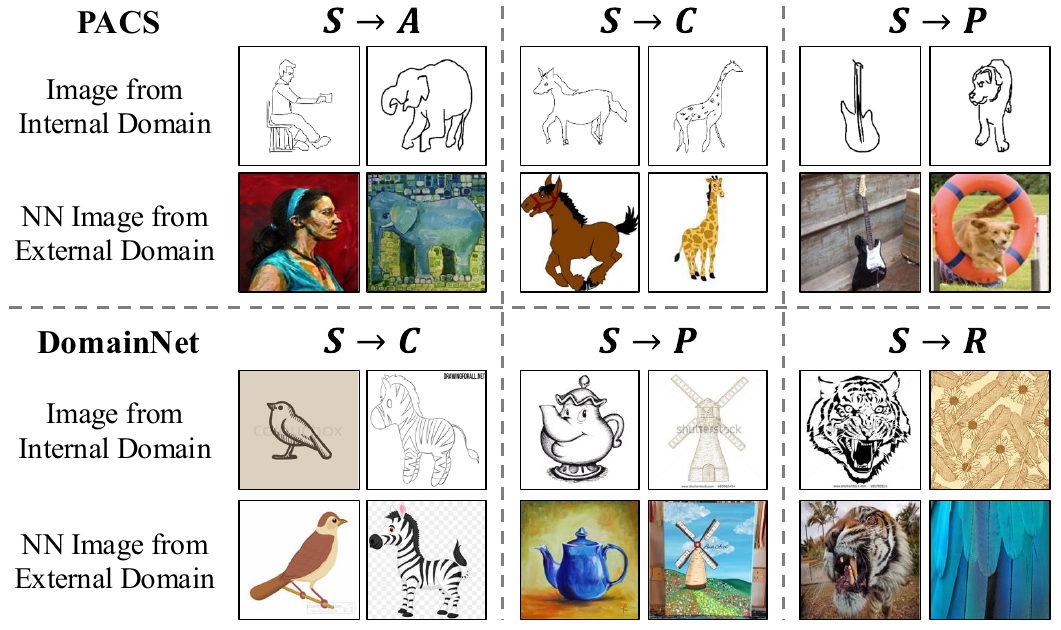}
    \caption{Nearest Neighbors on PACS and DomainNet.}
    \label{fig:vi}
\end{figure}

\begin{table*}[t]

\renewcommand{\arraystretch}{1}
\centering
\setlength{\tabcolsep}{6pt}
{\small
\begin{tabular}{c||c|c|c|c|c|c||c|c|c|c|c}
\hline\hline

   \rowcolor{gray!20}
   \hline
 Variant &$P^G$   &   $P^D$&$\mathcal{L}_{\text{Con}}$&$DPG$   &  $MST$    & $t_{target}$ &  A   &   C  &  P&S&Average   \\
  \hline
\circnum[\small]{1}&\textcolor[rgb]{0.13, 0.55, 0.13}{\ding{51}} &  & && &  & 91.21&90.94&94.54&93.61&92.58\\ 
   \rowcolor{gray!10}
     \circnum[\small]{2}& & \textcolor[rgb]{0.13, 0.55, 0.13}{\ding{51}} &&\textcolor[rgb]{0.13, 0.55, 0.13}{\ding{51}}& & &85.48&83.16&90.79&91.25&87.67\\ 

 \circnum[\small]{3}&\textcolor[rgb]{0.13, 0.55, 0.13}{\ding{51}} &  &&&\textcolor[rgb]{0.13, 0.55, 0.13}{\ding{51}} & &92.43&94.14&95.81&95.13&94.38\\ 
    \rowcolor{gray!10}
\circnum[\small]{4}&\textcolor[rgb]{0.13, 0.55, 0.13}{\ding{51}} &\textcolor[rgb]{0.13, 0.55, 0.13}{\ding{51}}  &\textcolor[rgb]{0.13, 0.55, 0.13}{\ding{51}}&\textcolor[rgb]{0.13, 0.55, 0.13}{\ding{51}}&  &  &96.03&97.44&98.98&95.77&97.06 \\ 

\circnum[\small]{5}&\textcolor[rgb]{0.13, 0.55, 0.13}{\ding{51}} &\textcolor[rgb]{0.13, 0.55, 0.13}{\ding{51}} &&\textcolor[rgb]{0.13, 0.55, 0.13}{\ding{51}}  & \textcolor[rgb]{0.13, 0.55, 0.13}{\ding{51}} &  & 97.74&99.09&99.48&95.48&97.95\\ 
   \rowcolor{gray!10}
   
  \circnum[\small]{6}& \textcolor[rgb]{0.13, 0.55, 0.13}{\ding{51}} &\textcolor[rgb]{0.13, 0.55, 0.13}{\ding{51}}&\textcolor[rgb]{0.13, 0.55, 0.13}{\ding{51}}& \textcolor[rgb]{0.13, 0.55, 0.13}{\ding{51}} &  \textcolor[rgb]{0.13, 0.55, 0.13}{\ding{51}} &  \textcolor[rgb]{0.13, 0.55, 0.13}{\ding{51}}
 & 97.93&99.68&99.85&96.27&98.42\\ 
 \hline
FaST-PT&\textcolor[rgb]{0.13, 0.55, 0.13}{\ding{51}} &\textcolor[rgb]{0.13, 0.55, 0.13}{\ding{51}}&\textcolor[rgb]{0.13, 0.55, 0.13}{\ding{51}}& \textcolor[rgb]{0.13, 0.55, 0.13}{\ding{51}} &  \textcolor[rgb]{0.13, 0.55, 0.13}{\ding{51}} &   
 & 97.32&99.31&99.93&96.09&98.16\\ 
    \hline\hline

\end{tabular}
}

\caption{Effect of different components of Fast-PT on PACS.}
\label{tab:ablation}
\end{table*}

\paragraph{Visualization of MST}
We evaluate the style transfer effect of the MST strategy, with the results presented in Figure~\ref{fig:vi}.
Since existing methods do not support converting image features back into images, we adopt a Nearest Neighbors (NN) approach to assess the generated augmented features. Specifically, we use all data from the extended domain as the reference set. For each enhanced embedding, we compute its cosine similarity with embeddings in the reference set and select the one with the highest similarity as its NN.
In Figure~\ref{fig:vi}, the top row displays the original images used for style transfer, while the bottom row shows the NNs corresponding to their enhanced features. We observe that although there is a clear difference in style between the top and bottom images, they consistently belong to the same semantic category. This observation aligns with the objectives of the two-part loss function employed during training and demonstrates the effectiveness of the MST strategy.

\begin{figure}[t]
    \centering
    \includegraphics[width=1.0\linewidth]{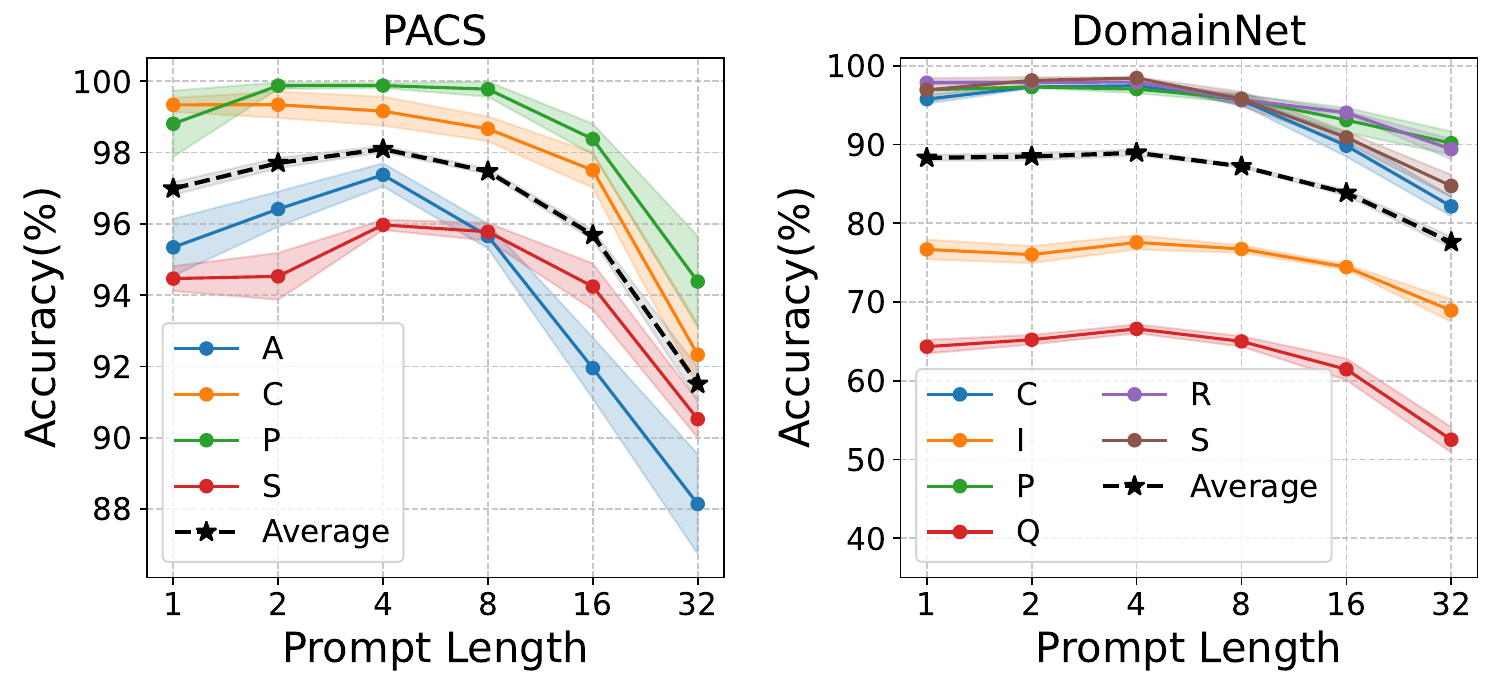}
    \caption{Effect of prompt length on PACS and DomainNet datasets over three runs with different random seeds. }
    \label{fig:len}
\end{figure}

\paragraph{Effect of prompt length}
We analyze the effect of prompt length and demonstrate the result in Figure  \ref{fig:len}. 
Here, we set the lengths of the global and domain prompts to be equal.
On PACS and DomainNet datasets, we observed that in most tasks, model performance exhibited a trend of initially increasing and then decreasing as the prompt length grows.
Short prompt constrains the performance due to the limited information conveyed in the prompt. Conversely, long prompts may introduce noise or task-irrelevant information, which can negatively impact the alignment between image and text. Moreover, when using long prompts, the prompt weighting strategy employed in FaST-PT may lead to issues such as semantic dilution and reduced interpretability. 
We observe that when the prompt length is set to 4, most tasks, as well as the average accuracy, achieve their optimal performance. Therefore, we adopt a prompt length of 4 as the default setting in our experiments.

\begin{figure}[t]
    \centering
    \includegraphics[width=1\linewidth]{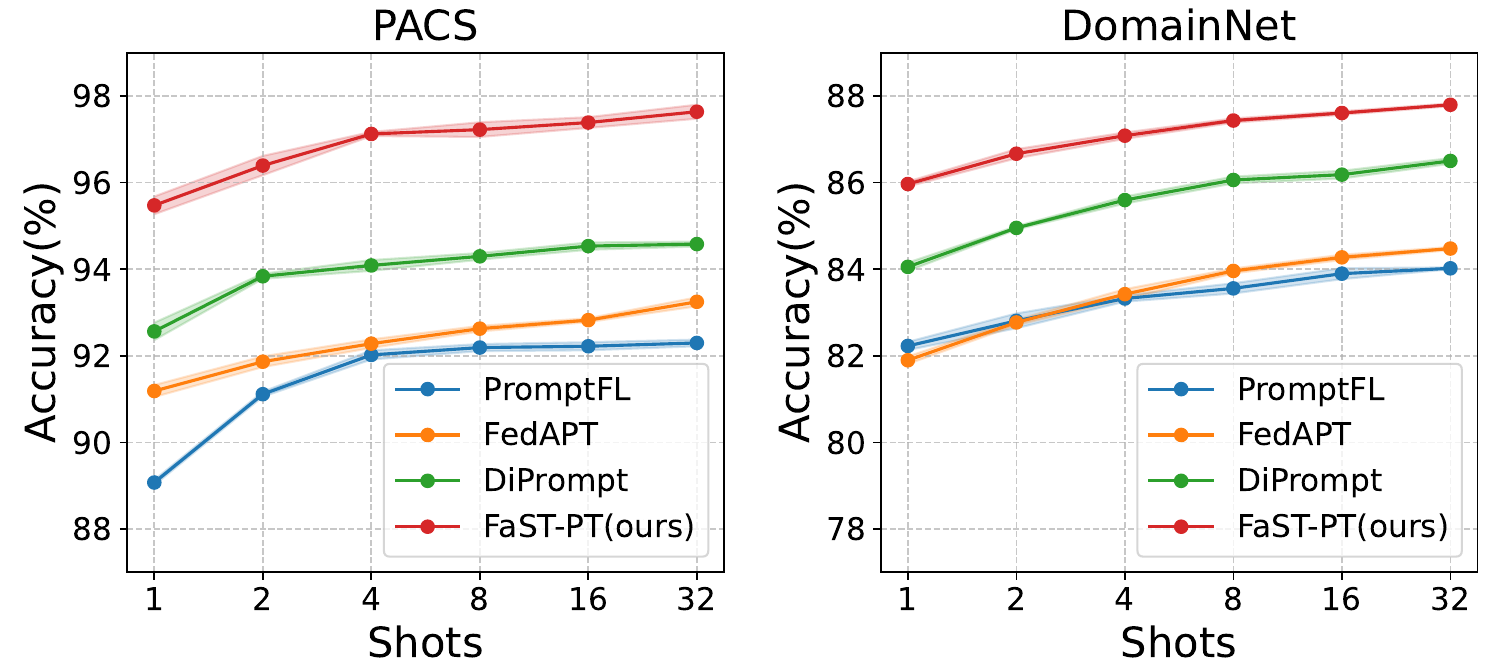}
    \caption{Few-shot performance on PACS and DomainNet datasets over three runs with different random seeds.}
    \label{fig:shot}
\end{figure}

\paragraph{Result under Few-shot scenario}
To investigate the effectiveness of our framework in a few-shot setting, we compare the generalization ability of FaST-PT with existing prompt-based methods under varying conditions of extremely limited data (shots = 1, 2, 4, 8, 16, 32). As illustrated in Figure \ref{fig:shot}, we observe that the performance of all methods improves with the increase in training samples. FaST-PT consistently demonstrates significant advantages across all settings on both the PACS and DomainNet datasets. These experimental results indicate that FaST-PT is capable of capturing both global and domain-specific information under few-shot settings, achieving robust generalization to unseen domains.

\paragraph{Cost Analysis}
We analysis the efficiency of FaST-PT in terms of communication and computation cost. We selected one representative method from each of the three categories of baseline approaches for analysis. For fairness, all methods employ ViT-L/14 as the visual encoder, with a consistent batch size of 16 and prompt length set to 4.
Regarding communication efficiency, we measure params each client uploads per training round. Compared to traditional FL that requires uploading the entire model, prompt-based FL methods incur significantly lower communication costs. Moreover, relative to DiPrompt, our method utilizes more lightweight and efficient modules, resulting in fewer than half the number of transmitted parameters.
In terms of computational cost, we evaluated the FLOPs for processing a batch during inference. The additional modules introduced by FedADG substantially increase inference cost, whereas our approach only adds a domain classifier. Consequently, its FLOPs are only slightly higher than FedAvg and remain lower than those of DiPrompt.

\begin{table}[t]

\renewcommand{\arraystretch}{1}
\setlength{\tabcolsep}{5pt}
\centering
 {\small
\begin{tabular}{c|c|c|c|c}
\hline\hline

Method & FedAvg & FedADG & DiPrompt & Ours \\
  \hline
 Parameters & $305M$ & $318M$ & $0.012M$&$0.005M$\\
 \hline
 FLOPs &$1.296T$&$2.593T$&$1.739T$&$1.431T$\\
\hline\hline 
\end{tabular}}

 \caption{Communication and computation cost compared with baseline FDG methods. ($Batch size = 16$)}
 \label{tab:commu}
\end{table}

\section{Conclusion}
In this work, we propose a new framework named FaST-PT, which integrates CLIP into FL to tackle the FDG problem in terms of local feature augmentation and efficient unseen domain adaptation.
Specifically, FaST-PT utilizes the MST strategy that allows clients to generate image embedding of external domains for mitigating the domain shift among clients. 
Besides, we design a prompt tuning module that partitions the prompt for capturing general and domain-specific knowledge, along with a domain-aware prompt generation mechanism to achieve the similarity-based domain knowledge mixing for efficient adaptation to unseen domains.
Extensive experiments demonstrate that FaST-PT outperforms existing FDG methods across all the generalization tasks on four benchmark cross-domain datasets.

\section{Acknowledgments}

This work was supported by the National Natural Science Foundation of China under Grant 62572311, 62202302, U21B2019, 62406347, and 62572260.

\bibliography{aaai2026}

@inproceedings{mcmahan2017communication,
  title={Communication-efficient learning of deep networks from decentralized data},
  author={McMahan, Brendan and Moore, Eider and Ramage, Daniel and Hampson, Seth and y Arcas, Blaise Aguera},
  booktitle={Artificial intelligence and statistics},
  pages={1273--1282},
  year={2017},
  organization={PMLR}
}

@article{konevcny2016federated,
  title={Federated learning: Strategies for improving communication efficiency},
  author={Kone{\v{c}}n{\`y}, Jakub and McMahan, H Brendan and Yu, Felix X and Richt{\'a}rik, Peter and Suresh, Ananda Theertha and Bacon, Dave},
  journal={arXiv preprint arXiv:1610.05492},
  year={2016}
}

@article{li2021fedbn,
  title={Fedbn: Federated learning on non-iid features via local batch normalization},
  author={Li, Xiaoxiao and Jiang, Meirui and Zhang, Xiaofei and Kamp, Michael and Dou, Qi},
  journal={arXiv preprint arXiv:2102.07623},
  year={2021}
}

@article{li2020federated,
  title={Federated optimization in heterogeneous networks},
  author={Li, Tian and Sahu, Anit Kumar and Zaheer, Manzil and Sanjabi, Maziar and Talwalkar, Ameet and Smith, Virginia},
  journal={Proceedings of Machine learning and systems},
  volume={2},
  pages={429--450},
  year={2020}
}

@article{lester2021power,
  title={The power of scale for parameter-efficient prompt tuning},
  author={Lester, Brian and Al-Rfou, Rami and Constant, Noah},
  journal={arXiv preprint arXiv:2104.08691},
  year={2021}
}

@inproceedings{radford2021learning,
  title={Learning transferable visual models from natural language supervision},
  author={Radford, Alec and Kim, Jong Wook and Hallacy, Chris and Ramesh, Aditya and Goh, Gabriel and Agarwal, Sandhini and Sastry, Girish and Askell, Amanda and Mishkin, Pamela and Clark, Jack and others},
  booktitle={International conference on machine learning},
  pages={8748--8763},
  year={2021},
  organization={PMLR}
}

@inproceedings{zhou2022conditional,
  title={Conditional prompt learning for vision-language models},
  author={Zhou, Kaiyang and Yang, Jingkang and Loy, Chen Change and Liu, Ziwei},
  booktitle={Proceedings of the IEEE/CVF Conference on Computer Vision and Pattern Recognition},
  pages={16816--16825},
  year={2022}
}

@inproceedings{yang2023efficient,
  title={Efficient model personalization in federated learning via client-specific prompt generation},
  author={Yang, Fu-En and Wang, Chien-Yi and Wang, Yu-Chiang Frank},
  booktitle={Proceedings of the IEEE/CVF International Conference on Computer Vision},
  pages={19159--19168},
  year={2023}
}

@inproceedings{su2024federated,
  title={Federated adaptive prompt tuning for multi-domain collaborative learning},
  author={Su, Shangchao and Yang, Mingzhao and Li, Bin and Xue, Xiangyang},
  booktitle={Proceedings of the AAAI Conference on Artificial Intelligence},
  volume={38},
  number={13},
  pages={15117--15125},
  year={2024}
}

@article{guo2023promptfl,
  title={Promptfl: Let federated participants cooperatively learn prompts instead of models-federated learning in age of foundation model},
  author={Guo, Tao and Guo, Song and Wang, Junxiao and Tang, Xueyang and Xu, Wenchao},
  journal={IEEE Transactions on Mobile Computing},
  year={2023},
  publisher={IEEE}
}

@inproceedings{zhao2023fedprompt,
  title={FedPrompt: Communication-Efficient and Privacy-Preserving Prompt Tuning in Federated Learning},
  author={Zhao, Haodong and Du, Wei and Li, Fangqi and Li, Peixuan and Liu, Gongshen},
  booktitle={ICASSP 2023-2023 IEEE International Conference on Acoustics, Speech and Signal Processing (ICASSP)},
  pages={1--5},
  year={2023},
  organization={IEEE}
}

@inproceedings{peng2019moment,
  title={Moment matching for multi-source domain adaptation},
  author={Peng, Xingchao and Bai, Qinxun and Xia, Xide and Huang, Zijun and Saenko, Kate and Wang, Bo},
  booktitle={Proceedings of the IEEE/CVF international conference on computer vision},
  pages={1406--1415},
  year={2019}
}

@inproceedings{wu2021collaborative,
  title={Collaborative optimization and aggregation for decentralized domain generalization and adaptation},
  author={Wu, Guile and Gong, Shaogang},
  booktitle={Proceedings of the IEEE/CVF international conference on computer vision},
  pages={6484--6493},
  year={2021}
}

@inproceedings{liu2021feddg,
  title={Feddg: Federated domain generalization on medical image segmentation via episodic learning in continuous frequency space},
  author={Liu, Quande and Chen, Cheng and Qin, Jing and Dou, Qi and Heng, Pheng-Ann},
  booktitle={Proceedings of the IEEE/CVF conference on computer vision and pattern recognition},
  pages={1013--1023},
  year={2021}
}

@article{nguyen2022fedsr,
  title={Fedsr: A simple and effective domain generalization method for federated learning},
  author={Nguyen, A Tuan and Torr, Philip and Lim, Ser Nam},
  journal={Advances in Neural Information Processing Systems},
  volume={35},
  pages={38831--38843},
  year={2022}
}

@inproceedings{le2024efficiently,
  title={Efficiently assemble normalization layers and regularization for federated domain generalization},
  author={Le, Khiem and Ho, Long and Do, Cuong and Le-Phuoc, Danh and Wong, Kok-Seng},
  booktitle={Proceedings of the IEEE/CVF Conference on Computer Vision and Pattern Recognition},
  pages={6027--6036},
  year={2024}
}

@inproceedings{chen2023federated,
  title={Federated domain generalization for image recognition via cross-client style transfer},
  author={Chen, Junming and Jiang, Meirui and Dou, Qi and Chen, Qifeng},
  booktitle={Proceedings of the IEEE/CVF Winter Conference on Applications of Computer Vision},
  pages={361--370},
  year={2023}
}

@inproceedings{huang2023rethinking,
  title={Rethinking federated learning with domain shift: A prototype view},
  author={Huang, Wenke and Ye, Mang and Shi, Zekun and Li, He and Du, Bo},
  booktitle={2023 IEEE/CVF Conference on Computer Vision and Pattern Recognition (CVPR)},
  pages={16312--16322},
  year={2023},
  organization={IEEE}
}

@inproceedings{muandet2013domain,
  title={Domain generalization via invariant feature representation},
  author={Muandet, Krikamol and Balduzzi, David and Sch{\"o}lkopf, Bernhard},
  booktitle={International conference on machine learning},
  pages={10--18},
  year={2013},
  organization={PMLR}
}

@inproceedings{li2018domain,
  title={Domain generalization via conditional invariant representations},
  author={Li, Ya and Gong, Mingming and Tian, Xinmei and Liu, Tongliang and Tao, Dacheng},
  booktitle={Proceedings of the AAAI conference on artificial intelligence},
  volume={32},
  number={1},
  year={2018}
}

@article{zhang2021federated,
  title={Federated learning with domain generalization},
  author={Zhang, Liling and Lei, Xinyu and Shi, Yichun and Huang, Hongyu and Chen, Chao},
  journal={arXiv preprint arXiv:2111.10487},
  year={2021}
}

@article{zhou2022learning,
  title={Learning to prompt for vision-language models},
  author={Zhou, Kaiyang and Yang, Jingkang and Loy, Chen Change and Liu, Ziwei},
  journal={International Journal of Computer Vision},
  volume={130},
  number={9},
  pages={2337--2348},
  year={2022},
  publisher={Springer}
}

@inproceedings{khattak2023maple,
  title={Maple: Multi-modal prompt learning},
  author={Khattak, Muhammad Uzair and Rasheed, Hanoona and Maaz, Muhammad and Khan, Salman and Khan, Fahad Shahbaz},
  booktitle={Proceedings of the IEEE/CVF conference on computer vision and pattern recognition},
  pages={19113--19122},
  year={2023}
}

@inproceedings{venkateswara2017deep,
  title={Deep hashing network for unsupervised domain adaptation},
  author={Venkateswara, Hemanth and Eusebio, Jose and Chakraborty, Shayok and Panchanathan, Sethuraman},
  booktitle={Proceedings of the IEEE conference on computer vision and pattern recognition},
  pages={5018--5027},
  year={2017}
}

@inproceedings{fang2013unbiased,
  title={Unbiased metric learning: On the utilization of multiple datasets and web images for softening bias},
  author={Fang, Chen and Xu, Ye and Rockmore, Daniel N},
  booktitle={Proceedings of the IEEE international conference on computer vision},
  pages={1657--1664},
  year={2013}
}

@inproceedings{li2017deeper,
  title={Deeper, broader and artier domain generalization},
  author={Li, Da and Yang, Yongxin and Song, Yi-Zhe and Hospedales, Timothy M},
  booktitle={Proceedings of the IEEE international conference on computer vision},
  pages={5542--5550},
  year={2017}
}

@article{bui2021exploiting,
  title={Exploiting domain-specific features to enhance domain generalization},
  author={Bui, Manh-Ha and Tran, Toan and Tran, Anh and Phung, Dinh},
  journal={Advances in Neural Information Processing Systems},
  volume={34},
  pages={21189--21201},
  year={2021}
}

@inproceedings{zhang2022towards,
  title={Towards principled disentanglement for domain generalization},
  author={Zhang, Hanlin and Zhang, Yi-Fan and Liu, Weiyang and Weller, Adrian and Sch{\"o}lkopf, Bernhard and Xing, Eric P},
  booktitle={Proceedings of the IEEE/CVF conference on computer vision and pattern recognition},
  pages={8024--8034},
  year={2022}
}

@article{zhang2021federated2,
  title={Federated learning with domain generalization},
  author={Zhang, Liling and Lei, Xinyu and Shi, Yichun and Huang, Hongyu and Chen, Chao},
  journal={arXiv preprint arXiv:2111.10487},
  year={2021}
}

@inproceedings{bai2024diprompt,
  title={Diprompt: Disentangled prompt tuning for multiple latent domain generalization in federated learning},
  author={Bai, Sikai and Zhang, Jie and Guo, Song and Li, Shuaicheng and Guo, Jingcai and Hou, Jun and Han, Tao and Lu, Xiaocheng},
  booktitle={Proceedings of the IEEE/CVF Conference on Computer Vision and Pattern Recognition},
  pages={27284--27293},
  year={2024}
}

@article{guan2024federated,
  title={Federated learning for medical image analysis: A survey},
  author={Guan, Hao and Yap, Pew-Thian and Bozoki, Andrea and Liu, Mingxia},
  journal={Pattern Recognition},
  pages={110424},
  year={2024},
  publisher={Elsevier}
}

\end{document}